\documentclass[aps,floatfix]{revtex4}
\usepackage{graphicx}
\usepackage{epsfig}

\begin{document}

\title{Scalar Polynomial Singularities in Power-Law Spacetimes}
\author{Kayll Lake\cite{email}}
\affiliation{Department of Physics and Department of Mathematics
and Statistics, Queen's University, Kingston, Ontario, Canada, K7L
3N6 }
\date{\today}

\begin{abstract}
Recently, Helliwell and Konkowski (\texttt{gr-qc/0701149}) have
examined the quantum ``healing" of some classical singularities in
certain power-law spacetimes. Here I further examine classical
properties of these spacetimes and show that some of them contain
naked strong curvature singularities.
\end{abstract}
\maketitle
\section{Power-law spacetimes}
The purpose of this study is to examine scalar polynomial
singularities associated with power-law spacetimes that can be
given in the form
\begin{equation}
ds^2=r^{\alpha}dt^2-r^{\beta}dr^2-\frac{r^{\gamma}}{C^2}d\theta^2-r^{\delta}(dz+Ad\theta)^2,\label{general}
\end{equation}
where $\alpha, \beta, \gamma$ and $\delta$ are (real) constants
and $A$ and $C$ are functions of $\theta$ only. Helliwell and
Konkowski \cite{hk} have recently considered the quantum
``healing" of classical singularities in the spacetimes
(\ref{general}) (for $C$ constant and $A=0$). An elementary
transformation in $r$ can be used to simplify (\ref{general}) into
two cases and in the notation of Helliwell and Konkowski these
are; $\alpha=\beta$ for $\alpha \neq \beta+2$ (Type I) and $\alpha
= \beta+2$ (Type II). The present discussion is organized as
follows: first we locate all polynomial singularities in these
spacetimes, then we examine the affine distance to the
singularities along timelike and null geodesics. Further, we
examine details in the $t-r$ subspaces, including the construction
of double-null coordinates and finally the focusing conditions
associated with the singularities.
\section{Singularities in Type I}
Consider the Newman-Penrose tetrad \cite{np}
\begin{equation}
l_{a}=r^{\delta}[\sqrt {{r}^{\beta-\delta}},0,-A,-1],\label{l1}
\end{equation}
\begin{equation}
n_{a}=\frac{1}{2}[\sqrt {{r}^{\beta-\delta}},0 ,A,1],\label{n1}
\end{equation}
\begin{equation}
m_{a}=\frac{1}{\sqrt{2}}[0,-\sqrt {{r}^{\beta}},-\frac {i \sqrt
{r^{\gamma}}}{C},0]\label{m1}
\end{equation}
and
\begin{equation}
\bar{m}_{{a}}=\frac{1}{\sqrt{2}}[0,-\sqrt {{r}^{\beta}},\frac {i
\sqrt {r^{\gamma}}}{C},0].\label{mbar1}
\end{equation}
This tetrad generates the Type I spacetime
\begin{equation} \label{type1ds}
ds^2=r^{\beta}(dt^2-dr^2)-\frac{r^{\gamma}}{C^2}d\theta^2-r^{\delta}(dz+Ad\theta)^2.
\end{equation}
The non-vanishing tetrad components of the trace-free Ricci tensor
are given by
\begin{equation}
\Phi 00={\frac {{r}^{\delta} \left( \delta+\gamma-2
 \right)  \left( \beta-\delta \right) }{8{r}^{\beta}{r}^{2}}},
\end{equation}
\begin{equation}
\Phi 02=\frac
{\beta\,\gamma-{\delta}^{2}+\delta\,\gamma+\beta\,\delta+2\,\beta+2\,\delta}{16{r}^{\beta}{r}^{2}},
\end{equation}
\begin{equation}
\Phi 11=\frac {-{\gamma}^{2}+\beta\,\gamma+\beta
\,\delta+2\,\gamma}{16{r}^{\beta}{r}^{2}}
\end{equation}
and
\begin{equation}
\Phi 22=\frac { ( \beta-\delta )  ( \delta+ \gamma-2)
}{32{r}^{\beta}{r}^{\delta}{r}^{2}}.
\end{equation}
The non-vanishing tetrad components of the Weyl tensor are given
by
\begin{equation}
\Psi 0=-\frac {{r}^{\delta} ( -\delta+\gamma+2
 )  ( \beta-\delta) }{8{r}^{\beta}{r}^{2}},
\end{equation}
\begin{equation}
\Psi 2=-\frac
{-3\,\beta\,\delta+{\delta}^{2}+3\,\beta\,\gamma+\delta\,\gamma-2\,{\gamma}^{2}+4\,\gamma-2\,
\beta-2\,\delta}{48{r}^{\beta}{r}^{2}}
\end{equation}
and
\begin{equation}
\Psi 4=-\frac { \left( -\delta+\gamma+2 \right)
 \left( \beta-\delta \right) }{32{r}^{\beta}{r}^{\delta}{r}^{2}}.
\end{equation}
Finally, the Ricci scalar is given by
\begin{equation}
R=-\frac {-\delta\,\gamma-{\delta}^{2}+2\,
\delta+2\,\gamma+2\,\beta-{\gamma}^{2}}{2{r}^{\beta}{r}^{2}}.
\label{Ricci}
\end{equation}

\bigskip

In general these spacetimes are of Petrov type I but reduce to
Petrov type D (and then O) for the following specializations:
$\gamma=\delta$ ($\beta=\delta$), $\gamma=\delta+2$
($\beta=\delta$), $\gamma=\beta$ ($\beta=\delta$,
$\beta=\delta-2$), $\gamma=2\beta+2-\delta$  ($\beta=\delta$,
$\beta=\delta-2$).

\bigskip

The tetrad components of the trace-free Ricci and Weyl tensors
vanish identically in four cases: (i)
 $(\beta=\gamma=\delta=0)$, (ii) $(\beta=\delta=0, \gamma=2)$, (iii)
 $(\beta=\gamma=0,\delta=2)$ and (iv) $(\beta=-2,
 \delta=\gamma=-2)$. In cases (i), (ii) and (iii) the spacetime is
flat. In case (iv) the Ricci scalar reduces to $12$.

\bigskip

Invariants of any spacetime consist of the Ricci scalar, the
invariant of lowest degree, and invariants of higher degree
constructed from appropriate products of the tetrad components of
the trace-free Ricci and Weyl tensors \cite{CM} \cite{Kret}. From
(\ref{Ricci}) it follows that $R$ diverges like
$1/{r}^{\beta}{r}^{2}$ as $r\rightarrow0$ unless
\begin{equation}
\beta=\gamma^2/2-\gamma+\delta^2/2-\delta+\delta \gamma/2.
\label{beta}
\end{equation}
Substitution of $\beta$ from (\ref{beta}) into the Ricci invariant
of next degree shows that the invariant diverges like
$(1/{r}^{\beta}{r}^{2})^2$ as $r\rightarrow0$ unless \cite{unless}
\begin{equation}
(\delta-2+\gamma)^2(\delta^2-2 \delta+4+\gamma \delta-2
\gamma+\gamma^2)=0.\label{next}
\end{equation}
Substitution of $\beta$ from (\ref{beta}) and $\delta=2-\gamma$
from (\ref{next}) into the first Weyl invariant shows that this
invariant diverges like $(1/{r}^{\beta}{r}^{2})^2$ as
$r\rightarrow0$ unless
\begin{equation}
\gamma^2(\gamma-2)^2=0.
\end{equation}
The case $\gamma=0$ gives $\delta=2$ and $\beta=0$, that is, the
flat case (iii). The case $\gamma=2$ gives $\delta=0$ and
$\beta=0$, that is the flat case (ii). Substitution of $\beta$
from (\ref{beta}) and $\delta^2-2 \delta+4+\gamma \delta-2
\gamma+\gamma^2=0$ from (\ref{next}) into the first Weyl invariant
shows that this invariant diverges as $(1/{r}^{\beta}{r}^{2})^2$
as $r\rightarrow0$ for all real $\gamma$ and $\delta$. Finally,
note that both factors in (\ref{next}) cannot vanish
simultaneously for real $\gamma$ and $\delta$.

\bigskip

It follows that for Type I power-law spacetimes, except for the
four cases (i) through (iv), the spacetimes contain a scalar
polynomial singularity at $r=0$ for all $\beta+2>0$.

\section{Singularities in Type II}
Now consider the Newman-Penrose tetrad consisting of (\ref{l1})
and (\ref{n1}), both with $\beta$ replaced by $\beta+2$, along
with (\ref{m1}) and (\ref{mbar1}) unchanged. This tetrad generates
the Type II spacetime
\begin{equation}\label{type2ds}
ds^2=r^{\beta+2}(dt^2-\frac{dr^2}{r^2})-\frac{r^{\gamma}}{C^2}d\theta^2-r^{\delta}(dz+Ad\theta)^2.
\end{equation}
The non-vanishing tetrad components of the trace-free Ricci tensor
are now given by
\begin{equation}
\Phi 00={\frac {{r}^{\delta} \left(\beta-\delta+2
 \right)  \left( \delta+\gamma \right) }{8{r}^{\beta}{r}^{2}}},
\end{equation}
\begin{equation}
\Phi 02=\frac
{\beta\,\gamma-{\delta}^{2}+\delta\,\gamma+\beta\,\delta+2\,\gamma+2\,\delta}{16{r}^{\beta}{r}^{2}},
\end{equation}
\begin{equation}
\Phi 11=\frac {-{\gamma}^{2}+\beta\,\gamma+\beta
\,\delta+2\,\gamma+2\,\delta}{16{r}^{\beta}{r}^{2}}
\end{equation}
and
\begin{equation}
\Phi 22=\frac { ( \beta-\delta +2)  ( \delta+ \gamma)
}{32{r}^{\beta}{r}^{\delta}{r}^{2}}.
\end{equation}
The non-vanishing tetrad components of the Weyl tensor are now
given by
\begin{equation}
\Psi 0=-\frac {{r}^{\delta} ( -\delta+\beta+2
 )  ( \gamma-\delta) }{8{r}^{\beta}{r}^{2}},
\end{equation}
\begin{equation}
\Psi 2=-\frac {\left( \gamma-\delta \right)
\left(-\delta-2\gamma+6+3\beta \right) }{48{r}^{\beta}{r}^{2}},
\end{equation}
and
\begin{equation}
\Psi 4=-\frac { \left( -\delta+\beta+2 \right)
 \left( \gamma-\delta \right) }{32{r}^{\beta}{r}^{\delta}{r}^{2}}.
\end{equation}
Finally, the Ricci scalar is now given by
\begin{equation}
R=\frac {{\delta}^{2}+\,
\delta\gamma+\,{\gamma}^{2}}{2{r}^{\beta}{r}^{2}}. \label{Ricci2}
\end{equation}

\bigskip

In general these spacetimes are of Petrov type I but reduce to
Petrov type O for $\gamma=\delta$, and Petrov type D (and then O)
for the following specializations: $\gamma=\beta+2$
($\beta=\delta-2$), $\gamma=2\beta+4-\delta$ ($\beta=\delta-2$).

\bigskip

The tetrad components of the trace-free Ricci and Weyl tensors now
vanish identically only for $\delta=\gamma=0$ in which case the
spacetime is flat. Except in this case the Ricci scalar itself
diverges like $1/{r}^{\beta}{r}^{2}$ as $r\rightarrow0$.

\bigskip

It follows that for Type II power-law spacetimes, except for the
flat case $\delta=\gamma=0$, the spacetimes contain a scalar
polynomial singularity at $r=0$ for all $\beta+2>0$.

\section{Affine distance to the singularities}

Geodesics of the spacetimes (\ref{general}) satisfy
\begin{equation}
r^{\beta}\dot{r}^2=\frac{c_1^2}{r^{\alpha}}-\frac{r^{\gamma}\dot{\theta}^2}{C^2}-\frac{c_2^2}{r^{\delta}}-2\mathcal{L}
\label{geodesics}
\end{equation}
where $2\mathcal{L}=0$ in the null case and $2\mathcal{L}=1$ in
the timelike case, $c_1$ and $c_2$ are constants of the motion and
$ \dot{} \equiv d/d\lambda$ where $\lambda$ is an affine
parameter. It follows from (\ref{geodesics}) that
\begin{equation}
\lambda_{*}- \lambda_{0} \geq
\int_{0}^{r_{*}}r^{(\alpha+\beta)/2}dr \label{affine}
\end{equation}
where $\lambda_{*}$ and $r_{*}$ are finite and non-zero. As a
result, the singularities at $r=0$ are at finite affine distance
if and only if
\begin{equation}
\alpha+\beta+2>0.\label{finite}
\end{equation}
Whereas condition (\ref{finite}) offers no further restriction on
$\beta$ for singularities in Type II spacetimes (we again have
$\beta+2>0$), for Type I spacetimes we obtain the refined
condition $\beta+1>0$ for singularities, assuming, as usual, that
they must be at a finite affine distance.

\section{the $t-r$ subspaces $\Sigma$}
In what follows we examine in further detail the $t-r$ subspaces
(designated by $\Sigma$) of the spacetimes (\ref{general}). Since
we are primarily interested in the structure of the singularities
at $r=0$ we first explore the conditions under which null
geodesics can terminate there. Now transforming from $\theta$ to
$\tilde{\theta}$ where $d \theta/C(\theta)=d \tilde{\theta}$ we
have a constant of the motion $c_{3}$ and the null geodesic
equation can be given as
\begin{equation}\label{fullnull}
r^{\beta}
\dot{r}^2=\frac{1}{r^{\alpha}}-\frac{c_{2}^2}{r^{\delta}}-\frac{c_{3}^2}{r^{\gamma}}
\end{equation}
where, without loss in generality, we have set $c_{1}^2=1$. There
are four cases of interest: (i) If $c_{2}=c_{3}=0$ then clearly
there are no turning points $\dot{r}=0$ at finite $r$. (ii) If
$c_{2}=0, c_{3}\neq0$ then there is a minimum $r$ at $r_{0}>0$ for
$\alpha \geq \gamma$. (iii) If $c_{2}=0, c_{3}\neq0$ then there is
a minimum $r$ at $r_{0}>0$ for $\alpha \geq \delta$. (iv) If
$c_{2} \neq 0$ and $c_{3}\neq0$ then there is a minimum $r$ at
$r_{0}$ where
\begin{equation}\label{min}
\frac{1}{r_{0}^{\alpha}}=\frac{c_{2}^2}{r_{0}^{\delta}}+\frac{c_{3}^3}{r_{0}^{\gamma}}
\end{equation}
as long as the constants of the motion satisfy
\begin{equation}\label{c2}
c_{2}^2(\delta-\gamma)\leq(\alpha-\gamma)r_{0}^{\delta-\alpha}
\end{equation}
and
\begin{equation}\label{c3}
c_{3}^2(\gamma-\delta)\leq(\alpha-\delta)r_{0}^{\gamma-\alpha}.
\end{equation}
If there are minima in $r$ along null geodesics then a discussion
in the $t-r$ subspace of the spacetimes (\ref{general}) below the
minima is adequate for these geodesics in the sense that they do
not populate that region of the $t-r$ subspace. However, since
there are geodesics for which $c_{2}$ and $c_{3}$ do not vanish
simultaneously and which terminate at $r=0$, the $t-r$ subspaces
are not totally geodesic and discussions restricted to these
subspaces are not complete. However, an examination of these
subspaces is, though incomplete, very instructive. We now
construct double- null coordinates and examine focusing
conditions.

\subsection{Double Null Coordinates}
It is clear from (\ref{type1ds}) and (\ref{type2ds}) that the
trajectories $r=r_{0}=constant >0$ are timelike for $r>0$.
However, the nature of the singularities at $r=0$ is not clear.
Here we construct double-null coordinates to clarify the nature of
the singularities. By double-null coordinates we mean coordinates
which label affinely parameterized null geodesics of the full
spacetime. The associated spacetime diagrams that we draw are,
however, restricted to the $t-r$ subspaces.
\subsubsection{Type I}
On $\Sigma$ we have
\begin{equation}\label{sigma1}
ds^2_{\Sigma}=r^{\beta}(dt-dr)(dt+dr).
\end{equation}
Introducing null coordinates $u$ and $v$ where $u=constant$ along
$dt=dr$ and $v=constant$ along $dt=-dr$ and writing $r=f(u,v)$ and
$t=g(u,v)$ it follows that
\begin{equation}\label{type1uv}
f(u,v)=g(u,v)+F(u)=-g(u,v)+H(v).
\end{equation}
With the choices
\begin{equation}\label{type1uvchoice}
F(u)=2u,\;\;\;H(v)=2v,
\end{equation}
we have
\begin{equation}\label{type1rt}
r=u+v,\;\;\;t=v-u,
\end{equation}
so that
\begin{equation}\label{dstype1uv}
ds^2_{\Sigma}=-4(u+v)^{\beta}dudv
\end{equation}
where we have orientated the future so that $dv>0$ but $du<0$. We
have the following tangents to null geodesics in the full
spacetime (\ref{type1ds}) subject to the transformations
(\ref{type1rt}): along $v=v_{0}=constant$
\begin{equation}\label{k^atype1}
k^{a}= -\frac{\delta_{u}^{a}}{(u+v_{0})^{\beta}}
\end{equation}
so that $k^{b}\nabla_{b}k^{a}=0$ and along $u=u_{0}=constant$
\begin{equation}\label{l^a}
l^{a}= \frac{\delta_{v}^{a}}{(v+u_{0})^{\beta}}
\end{equation}
so that $l^{b}\nabla_{b}l^{a}=0$. As a result, the singularities
at $r=0$ are at finite affine distance only for $\beta+1>0$ as
discussed above. The trajectories $r=constant$ are straight
vertical lines and timelike throughout the $u-v$ diagram.
Trajectories of constant $t$ are straight horizontal lines and
spacelike throughout the $u-v$ diagram. A diagram is shown below
in Figure\ref{type1f}.
\begin{figure}[ht]
\epsfig{file=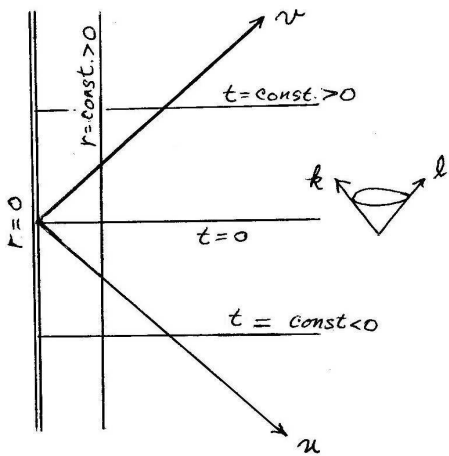,height=2.5in,width=2.5in,angle=0}
\caption{\label{type1f}Representation of the $t-r$ subspace for
Type I power-law spacetimes in the double-null coordinates $u$ and
$v$. The singularity at $r=0$ is timelike.}
\end{figure}
\subsubsection{Type II}
On $\Sigma$ we now have
\begin{equation}\label{sigma2}
ds^2_{\Sigma}=r^{\beta+2}(dt-\frac{dr}{r})(dt+\frac{dr}{r}).
\end{equation}
Introducing null coordinates $u$ and $v$ where now $u=constant$
along $dt=\frac{dr}{r}$ and $v=constant$ along $dt=-\frac{dr}{r}$
and again writing $r=f(u,v)$ and $t=g(u,v)$ it now follows that
\begin{equation}\label{type2uv}
f(u,v)=F(u)e^{g(u,v)}=H(v)e^{-g(u,v)}.
\end{equation}
With the choices
\begin{equation}\label{type2uvchoice}
F(u)=u^2,\;\;\;H(v)=v^2,
\end{equation}
so that $u$ and $v$ are not simultaneously zero, we have
\begin{equation}\label{type2rt}
r=uv,\;\;\;t=\ln(\frac{v}{u}),
\end{equation}
so that
\begin{equation}\label{nulltype2}
ds_{\Sigma}^{2}=-4 (u v)^{\beta +1}d u d v
\end{equation}
where again we have orientated the future so that $dv >0$ but $d u
< 0$. We now have tangents to null geodesics in the full spacetime
(\ref{type2ds}) subject to the transformations (\ref{type2rt})
\begin{equation}\label{k^a2}
k^{a}= -\frac{\delta_{u}^{a}}{u^{\beta+1}}
\end{equation}
so that $k^{b}\nabla_{b}k^{a}=0$ and
\begin{equation}\label{l^a2}
l^{a}= \frac{\delta_{v}^{a}}{v^{\beta+1}}
\end{equation}
so that $l^{b}\nabla_{b}l^{a}=0$. As a result, the singularities
at $r=0$ are at finite affine distance only for $\beta+2>0$ also
as shown above. The surfaces $r=constant >0$ are now hyperbolae in
the $u-v$ diagram and timelike. They become the degenerate null
hyperbola for $r=0$. Trajectories of constant $t$ are straight
lines through the origin $u=v=0$ which is a point of internal
infinity. A diagram is shown below in Figure\ref{type2f}.
\begin{figure}[ht]
\epsfig{file=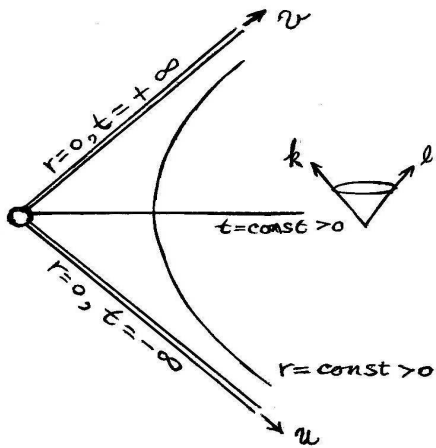,height=2.5in,width=2.5in,angle=0}
\caption{\label{type2f}Representation of the $t-r$ subspace for
Type II power-law spacetimes in the double-null coordinates $u$
and $v$. The singularity at $r=0$ is either past or future null.
The origin $u=v=0$ is a point of internal infinity.}
\end{figure}

\subsection{Focusing conditions} In this section we consider a
unified treatment of both types of spacetime. The vector
\begin{equation}
k^{a}=(1/r^{\alpha},\pm 1/r^{(\alpha+\beta)/2},0,0)\label{null}
\end{equation}
is tangent to a class of null geodesics of the spacetimes
(\ref{general}) and we use (\ref{null}) to examine focusing
conditions in the $t-r$ subspaces. Define
\begin{equation}
\Psi\equiv\lambda^nR_{ab}k^{a}k^{b}\label{focus}
\end{equation}
where $R_{ab}$ is the Ricci tensor. We find
\begin{equation}
\Psi=r^{(n/2-1)(\alpha+\beta+2)} \Delta\label{focuspower}
\end{equation}
where
\begin{equation}
\Delta=2\beta(\gamma+\delta)+\gamma(2-\gamma)+\delta(2-\delta)\equiv\Delta_{I}\label{deltaI}
\end{equation}
for Type I and
\begin{equation}
\Delta=2\beta(\gamma+\delta)+\gamma(4-\gamma)+\delta(4-\delta)\equiv\Delta_{II}\label{deltaII}
\end{equation}
for Type II. Clearly $\Psi=0 \;\forall \;n$ with $\Delta=0$. For
$\Delta\neq0$, $\Psi\rightarrow0$ as $r\rightarrow0$ for $n>2$ and
$|\Psi|\rightarrow\infty$ as $r\rightarrow 0$ for $n<2$. For
$n=2$,  $\Psi=\Delta$ and for $\Delta>0$ the singularities at
$r=0$ satisfy the strong curvature condition \cite{clarke}.
\section{Conclusions}
 Type I spacetimes can have strong curvature timelike
naked singularities at $r=0$ for $\beta+1>0$ and $\Delta_{I}>0$.
Type II spacetimes can have strong curvature past-null naked
singularities at $r=0$ for $\beta+2>0$ and $\Delta_{II}>0$.
\begin{acknowledgments}
It is a pleasure to thank Thomas Helliwell and Deborah Konkowski
for comments and in particular for suggesting a clarification of
the Type II singularities. This work was supported by a grant from
the Natural Sciences and Engineering Research Council of Canada
and was made possible by use of \textit{GRTensorII} \cite{grt}.
\end{acknowledgments}


\begin{thebibliography}{}\label{sec:TeXbooks}
\bibitem[*]{email}{Electronic Address: lake@astro.queensu.ca}
\bibitem{hk} T. M. Helliwell and D. A. Konkowski gr-qc/0701149
\bibitem{np}See, for example, H. Stephani, D. Kramer, M. MacCallum, C. Hoenselaers and
E. Herlt, \textit{Exact Solutions of Einstein's Field Equations}
(Cambridge University Press, Cambridge, 2003) chapter 7.
\bibitem{CM}See, for example, H. Stephani \textit{et al.} \cite{np} chapter 9.
\bibitem{Kret}The Kretschmann scalar (the full contraction of the
Riemann tensor) is not used here since it is a linear combination
of the square of the Ricci scalar, the first Ricci invarinat and
the real component of the first Weyl invariant.
\bibitem{unless}We exclude factors which have no non-vanishing real roots.
\bibitem{clarke} See, for example, C. J. S. Clarke,\textit{ The Analysis of Space-Time Singularities}
  (Cambridge University Press, Cambridge, 1993).
  \bibitem{grt}This is a package which runs within Maple. It is entirely
distinct from packages distributed with Maple and must be obtained
independently. The GRTensorII software and documentation is
distributed freely on the World-Wide-Web from the address \textit{
http://grtensor.org}
\end{thebibliography}
\end{document}